\documentclass[prb,twocolumn]{revtex4-1} 

\usepackage{amsmath}  
\usepackage{amsfonts} 
\usepackage{graphicx} 

\usepackage{braket}  
\usepackage{epstopdf}

\usepackage{xcolor} 
\newcommand{\bluee}{\textcolor{black}} 
\usepackage{ulem}   

\makeatother

\begin{document}

\title{A Wave Packet Approach to Resonant Scattering}
\author{A. M. Michalik and F. Marsiglio}
\address{Department of Physics, University of Alberta, Edmonton, Alberta, Canada,
T6G 2E1}

\date{\today}

\begin{abstract}
Resonant transmission occurs when constructive interference results
in the complete passage of an incoming wave through an array of barriers.
In this paper we explore such a scenario with one dimensional models.
We adopt wave packets with finite width to illustrate the deterioration
of resonance with decreasing wave packet width, and suggest an approximate
wave function for the transmitted and reflected components, derived
from aspects of both the wave packet and plane wave approaches. A comparison
with exact numerical calculations shows excellent agreement, and provides
insight into the scattering process. 
\end{abstract}
\maketitle

\section{Introduction\label{sec:introduction}}

Scattering experiments are one of the most effective ways of probing matter. Going back into the distant past,
physicists have been throwing projectiles at  objects to learn something about the projectile, the object, or
the interaction between the two. For example, Newton used a prism to learn about the properties of the projectile (light),
and Young used a double slit experiment for the same purpose.

Over the course of the last century, however, scattering experiments  have been  associated more with learning about the object. A
classic example is the series of alpha scattering experiments by Geiger and Marsden\citep{geiger1913} with guidance from Rutherford.
In this experiment, alpha particles were used to learn about the inner structure of the atom. Modern particle accelerators use higher energy
particle beams to probe the internal structure of the nucleus. 
Larger scale targets, be they molecules or solids, utilize the same principle with using 
X-rays or neutron beams. In each case the incoming energy can be varied, and the scattering profile can be measured as a function of angle and
outgoing energy. These are all three-dimensional problems, and our understanding of them is well developed
partly because we focus on the simplified scenario where the incoming and outgoing projectiles are well-described by a plane wave.
In this paper we move beyond the plane wave approximation and consider incoming and outgoing wave packets. To keep things relatively simple
we follow the traditional procedure used in the plane wave  approach by first restricting ourselves to one dimension. The two- or three-dimensional case is considerably
more complicated and is beyond the scope of this article.

Several recent descriptions of one-dimensional quantum scattering
have emphasized the wave packet approach. \citep{kim0673,kim0674,norsen08,schonhammer19,staelens21,phet}
 All of this work has espoused
the conceptual advantages of real-time wave packet scattering over
the stationary plane wave approximation found in standard undergraduate quantum texts,
and in addition, two of
these groups have adopted a lattice approach. For example, Ref.~\onlinecite{schonhammer19} emphasized the differences with the continuum approach and focused on narrow wave packets. In Ref.~\onlinecite{staelens21} one simulation focused
on a dimer (two-impurity) barrier where resonant transmission was known to take place
and modeled the time-dependent scattering interference. Our group\citep{kim0673,kim0674,staelens21} has exploited a remarkable feature: the lack of spreading for reasonably broad Gaussian wave packets on a lattice for certain wave vectors. 
Such wave packets maintain their qualitative characteristics during propagation and, therefore, make the effects of the scattering process clearer. 

In this work, we want to do two things. First, we will generalize
the possibilities for resonant transmission by considering any number
of barriers. Various aspects of this problem have already appeared
in the literature,\citep{griffiths92,sprung93} with some variation
of what we consider here, so we will merely provide a brief overview
of the theory. More specifically, we  consider  scattering
from a periodic array of $N$ identical impurities/barriers, each
spaced $m$ lattice sites apart. This problem is likely to be more
familiar to the reader in the continuum limit, and has been addressed
through the transfer matrix formalism in Ref.~\onlinecite{griffiths01}.
The tight-binding limit of various models of this problem on a lattice
has also been treated in Refs.~\onlinecite{kim0673,kim0674,markos08}.
For completeness, a derivation is presented in the online Supplementary Material. The
result of this derivation is that, given a barrier configuration consisting
of $N$ impurity potentials, each with strength $V$, spaced $m$
lattice spacings apart, then, for a particular dispersion relation,
there will exist a number of plane waves with fixed wave vector that
will yield unit transmission and zero reflectance.

However, there is a conceptual shortcoming with a time-{\it independent} plane wave approach: it does not give a complete qualitative understanding of the scattering process, especially during resonant transmission.
From a plane wave description, we can calculate the transmission of reflection coefficients, but the components themselves are always plane waves before and after the scattering process. At resonance, the reflected component is exactly zero. Using a Gaussian wave packet that retains its shape under propagation and nonresonant scattering,  we can show that the reflected wave packet profile at resonance is non-zero, contrary to the wave packet description. It is also no longer Gaussian. This reflected shape profile is derivable from a plane wave description of sufficiently 
broad wave packets.

Therefore our second goal in this paper is to study how the scattering profile changes
when a Gaussian wave-packet with real-space width $\alpha<\infty$ is used (as $\alpha \rightarrow \infty$ we approach the plane wave description).
In this case, even at resonance, the transmission will be less than unity and a reflected portion will appear.
We can  study the dependence of the reflected portion on the wave packet width. Numerically we find
a very peculiar shape for the reflected packet, while the transmitted portion remains essentially Gaussian. In addition 
we derive a closed approximate analytical formula which works
remarkably well, both in describing the reflection and transmission,
and in giving the detailed form of the scattered wave packet profile.
We will examine the validity and limitations of this approximation
by comparing it with exact numerical solutions. It will be clear that
when barriers with extended spatial structure are used, increasingly
wide wave packets are required to recover the plane wave results.

As mentioned earlier, when the scattering problem is defined on a lattice, wave packets with a specific centroid wave vector do not spread with time,\citep{kim0673,kim0674,staelens21} unlike the continuum case. The lattice model
 requires the use of the tight-binding formalism (see online Supplementary Materials for technical details), but
 allows for a much clearer description of the scattering process. 
Senior undergraduate students should be able to understand the content of this paper and follow the required numerical work. These students should also be able to reproduce most of the paper's content as a long-term project, assuming they have completed a numerical methods prerequisite course.

\section{Plane-wave scattering on a lattice from a finite periodic array}

\label{sec:plane_wave}

Following Ref.~\onlinecite{staelens21} we adopt as a basis the site
representation, where, using bra-ket notation, each (orthonormal)
basis state, $|\ell\rangle\equiv c_{\ell}^{\dagger}|0\rangle$, represents
a particle at position $x_{\ell}\equiv\ell a$, with $a$ the lattice
spacing and $\ell$ an integer. Here, $c_{\ell}^{\dagger}$ ($c_{\ell}$)
represents the creation (annihilation) operator for a particle at
site $\ell$, and $\ket{0}$ is the vacuum state, i.e. the empty lattice
state, and $\ell$ is one of $L$ lattice sites. In principle, $L\rightarrow\infty$
while, in practice, $L$ is some large integer.  We also adopt periodic
boundary conditions.

Initially we imagine a plane wave impinging from the left; without
barriers this wave would continue unimpeded towards the right. Propagation
consists of ``hops'' from one site to another, and for simplicity
we consider here nearest-neighbour hops only. This part of the Hamiltonian
constitutes the kinetic energy part, denoted by $H_{0}$, and is written
as 
\begin{eqnarray}
H_{0} & = & -t_{0}\sum_{\ell}\left(c_{\ell}^{\dagger}c_{\ell+1}+c_{\ell+1}^{\dagger}c_{\ell}\right),\label{h_0}
\end{eqnarray}
where $t_{0}$ is the hopping amplitude. The eigenstates of this Hamiltonian
alone are Bloch states, with wave vector $ka\equiv2\pi n/L$ and
$n$ an integer with domain $-L/2<n\le L/2$.
\bluee{These eigenstates are given by
\begin{equation}
c_k^\dagger = {1 \over \sqrt{L}} \sum_{\ell^\prime} e^{ik\ell^\prime} c_{\ell^\prime}^\dagger,
\label{eigenstate}
\end{equation}
where the summation is over all sites. Application of the hopping Hamiltonian, Eq.~\ref{h_0}, to this state
confirms that it is an eigenstate with eigenvalue $E_k = -2t_0 \cos{(ka)}$.}

In general there is a finite barrier region, starting at site $I_{1}$.
This barrier consists of $N$ one-body potentials arranged periodically
with a barrier lattice spacing of $ma$, where $m$ is a positive
integer. The $N$ barriers are localized at sites $I_{1}$, $I_{2}\equiv I_{1}+m$,
$I_{3}\equiv I_{1}+2m$, ... $I_{N}\equiv I_{1}+(N-1)m$, and the
Hamiltonian describing the interaction with the incoming particle
(described for now as a plane wave) is 
\begin{eqnarray}
H_{V} & = & \sum_{\ell\in\mathcal{I}}V_{\ell}c_{\ell}^{\dagger}c_{\ell}=V\sum_{\ell\in\mathcal{I}}c_{\ell}^{\dagger}c_{\ell},\label{h_V}
\end{eqnarray}
where the second equality follows since we assume all barrier strengths
to be equal (hence the barrier has periodic structure). Since the
barrier is finite it has a total width given by 
\begin{equation}
w_{B}\equiv\left(N-1\right)ma.\label{barrier_width}
\end{equation}

A presumed solution for the wave function, 
\begin{eqnarray}
|\psi\rangle & = & \sum_{\ell=1}^{L}a_{\ell}|\ell\rangle\equiv\sum_{\ell=1}^{L}a_{\ell}c_{\ell}^{\dagger}|0\rangle,\label{psi}
\end{eqnarray}
can be substituted into the Schr\"odinger equation, $H|\psi\rangle=E|\psi\rangle$.
This procedure is described in the online Supplementary Materials, where appropriate plane wave
solutions are assumed to the left and right of the barrier region
(as is done in textbooks in the continuum limit). Using the transfer
matrix formalism (briefly reviewed in the online Supplementary Material), we ultimately
derive the simple result\citep{markos08} for the transmission $T$ on
a lattice with spacing $a$, of a plane wave with wave-vector $k$,
through $N$ barriers, each $m$ lattice spacings apart, and each
with strength $V$, 
\begin{equation}
T=\frac{1}{1+\left[v_{k}U_{N-1}\left(h_{k}\right)\right]^{2}},\label{transmission}
\end{equation}
where 
\begin{eqnarray}
v_{k} & \equiv & \frac{V}{2t_{0}\sin\left(ka\right)},\label{v_k}\\
h_{k} & \equiv & \cos\left(kma\right)+v_{k}\sin\left(kma\right),\label{h_k}
\end{eqnarray}
and $U_{n}(x)$ is the $n^{{\rm th}}$ Chebyshev polynomial of the
second kind (in $x$), with properties as described in the online Supplementary Material \ref{A:Proof}.
The reflection probability can easily be derived since the transmission
and reflection probabilities have to sum to unity, and therefore 
\begin{equation}
R=1-T=\frac{\left[v_{k}U_{N-1}\left(h_{k}\right)\right]^{2}}{1+\left[v_{k}U_{N-1}\left(h_{k}\right)\right]^{2}}\equiv\frac{f^{2}(k)}{1+f^{2}(k)},\label{reflection}
\end{equation}
where for convenience we define $f(k)\equiv v_{k}U_{N-1}(h_{k})$.
These quantities are further discussed in the online Supplementary Material, along with
the amplitudes $\rho(k)$ and $\tau(k)$, where $R\equiv|\rho|^{2}$
and $T\equiv|\tau|^{2}$. A version of Eq.~(\ref{transmission})
applicable to free electrons in the continuum seems to have been
first derived in Ref.~\onlinecite{griffiths92} for an array of $\delta$-functions,
and in Ref.~\onlinecite{sprung93} for a more general array;
see references therein for earlier related work. See also Ref.~\onlinecite{griffiths01}
for further discussion on transmission in the continuum limit.

\begin{figure}
\includegraphics{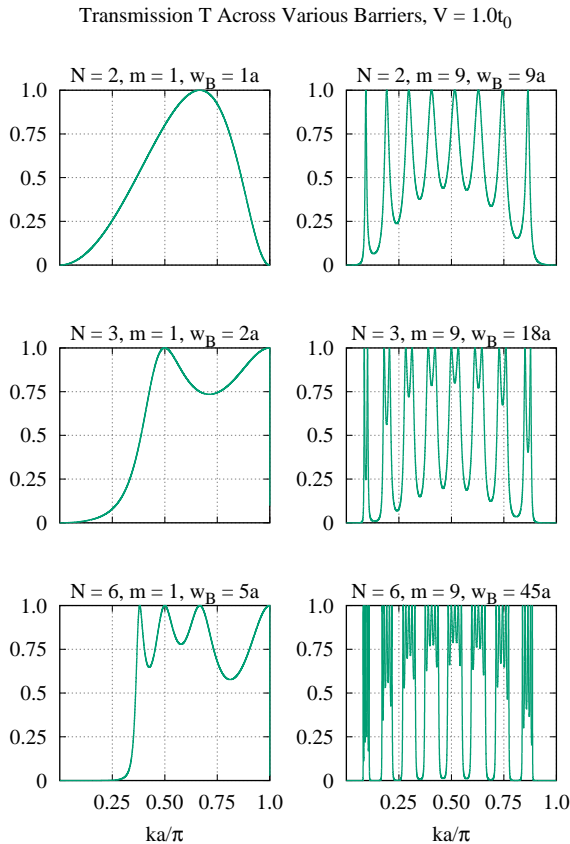}

\caption{\label{Fig1}Transmission $T$ vs. $ka/\pi$, across barriers with
strength $V=1.0t_{0}$ and various values of $N$, $m$ and $w_{B}$
as indicated above the various panels. Only positive values of $k$,
with range $0<ka/\pi<1$, are plotted. The energy, $E_{k}=-2t_{0}\cos(ka)$,
increases monotonically from $-2t_{0}$ to $+2t_{0}$ over this range.
Note the many instances of resonant transmission ($T=1$). As $V$
changes, the value of the wave vector where resonance will occur will
change according to Eq.~(\ref{resonance condition}). Note that $ka=\pi/2$
is special for wave packet propagation, as spreading with time is
limited at this wave vector, given the nearest-neighbor hopping model
we have adopted. Amongst these six examples unit transmission is clearly
seen to occur at this wave vector for all four cases beyond the first
row ($N=2$). }
\end{figure}

From the general transmission relation Eq.~(\ref{transmission})
we can calculate the barrier parameters for which unit transmission
(resonance) occurs. Unit transmission occurs \bluee{obviously} for $V=0$, i.e. no impurities,
or when $U_{N-1}(h_{k})=0$ for a non-zero value of $V$. The non-trivial
resonant condition is therefore 
\begin{eqnarray}
U_{N-1}\left(h_{k}\right) & = & 0,\label{resonance condition}
\end{eqnarray}
where $h_{k}=(h_{k}^{(n)})_{root}^{N-1}$ is the $n^{{\rm th}}$ root
of the $N-1$ Chebyshev polynomial, and the barrier parameters are
chosen such that 
\begin{eqnarray}
\left(h_{k}^{\left(n\right)}\right)_{root}^{N-1} & = & \cos\left(kma\right)+\frac{V\sin\left(kma\right)}{2t_{0}\sin\left(ka\right)}\label{h_k root}
\end{eqnarray}
is satisfied.

For a single impurity, $N=1$, only the trivial solution exists since
$U_{N-1}^{2}(h_{k})=U_{0}^{2}(h_{k})=1$ for all $k$, hence no resonance
can occur with a single impurity. For a barrier configuration with
$N\geq2$ impurities, a potential $V$ and wave vector $k$ can always
be found that satisfy Eq.~(\ref{h_k root}) for a given impurity
lattice spacing $ma$ and number of impurities $N$.

Examples of transmission across different barriers, of potential strength
$V=1.0\,t_{0}$, are shown in Fig.~\ref{Fig1} for selected values
of $N$ and $m$. It is clear from Eqs.~(\ref{resonance condition})-(\ref{h_k root}),
and as shown in Fig.~\ref{Fig1}, that resonant transmission ($T=1$)
is possible at many different values of the wave vector $k$, and
the number of these resonant points increases with increasing barrier
parameters $N$ and $m$. Plots similar to Fig.~\ref{Fig1} can be
found in many references, for example Refs.~\onlinecite{kim0673,kim0674,griffiths92,sprung93,griffiths01,markos08}.
\bluee{These plots all convey the idea that resonant transmission is possible at certain energies (i.e. values of wave vector k),
whose values depend on the specific parameters and details of the barriers. Students are taught that this is due to constructive
interference effects for a plane wave description. How is the resonant transmission and (lack of) reflection altered when wave packets 
are used? Before addressing this question we briefly review how wave packets are constructed on a lattice.}

\section{The wave packet description\label{sec:wave_packets}}

As mentioned in the introduction, there are many reasons to favor
a wave packet description of the incoming particle, not least of which is that
it allows a time-dependent evolution of the actual scattering event.
We follow previous prescriptions for such a wave packet description\citep{kim0673,kim0674,staelens21}
by constructing a set of eigenstates for the Hamiltonian that is the
sum of Eq.~(\ref{h_0}) and Eq.~(\ref{h_V}). Then the eigenstates
can be written formally as 
\begin{equation}
\ket{n}=\sum_{\ell}a_{\ell}^{\left(n\right)}c_{\ell}^{\dagger}\ket{0}\label{general_states}
\end{equation}
where $a_{\ell}^{(n)}$ are the eigenvector coefficients for the $n^{{\rm th}}$
eigenstate with eigenvalue $\epsilon_{n}$, both obtained through
numerical matrix diagonalization. It follows that the time-dependent
wave packet is 
\begin{equation}
\ket{\psi\left(t\right)}=\sum\limits _{n=1}^{L}\ket{n}\braket{n|\psi\left(0\right)}e^{-i\epsilon_{n}t/\hbar},\label{state_time}
\end{equation}
and the site-dependent wave function is given by $\langle\ell\ket{\psi(t)}$
to obtain 
\begin{equation}
\psi\left(x_{\ell},t\right)=\sum\limits _{n=1}^{L}a_{\ell}^{\left(n\right)}\braket{n|\psi(0)}e^{-i\epsilon_{n}t/\hbar},\label{wave_function_time}
\end{equation}
where $a_{\ell}^{(n)}$ represents the $\ell^{{\rm th}}$ element
of the $n^{{\rm th}}$ eigenvector and $\braket{n|\psi(0)}=\sum\nolimits _{\ell}[a_{\ell}^{(n)}]^{*}\phi(x_{\ell})$,
with the initial wave function given by 
\begin{equation}
|\psi\left(t=0\right)\rangle=\sum_{\ell}\phi\left(x_{\ell}\right)|\ell\rangle,\label{init}
\end{equation}
and the initial profile is described by a Gaussian, 
\begin{equation}
\phi\left(x_{\ell}\right)=\frac{1}{\left(2\pi\alpha^{2}\right)^{1/4}}e^{-\frac{1}{4}\left(x_{\ell}-x_{0}\right)^{2}/\alpha^{2}+ik_{0}\left(x_{\ell}-x_{0}\right)}.\label{free_space_wave_packet_initial}
\end{equation}
Here $x_{0}$ and $k_{0}$ are the mean position and mean wave vector
of the wave packet, respectively, and $\alpha$ is the initial uncertainty
in position (i.e. spread) of the wave packet. Once the eigenvalues
and eigenvectors are calculated, a straightforward summation in Eq.~(\ref{wave_function_time})
produces the wave function profile at any time thereafter. 

In fact this summation is relatively straightforward to implement in an empty
lattice (i..e no barriers whatsoever). It would give the novice student both the practice required to tackle the problem
with a barrier in place and the satisfaction of seeing a moving wave packet in motion [given by Eq.~(\ref{wave_function_time})]
that also does not spread if $k_0 a = \pi/2$ is used for the mean wave vector.\citep{cmp_remark}

One of the primary reasons for using wave packets specifically
defined on a lattice is that a wave vector can be found for which
the intrinsic spreading is essentially eliminated. 
A very good approximation to the time-dependent
wave function,  previously derived in
Refs.~\onlinecite{kim0673,kim0674} for an empty lattice, is given by 
\begin{align}
\psi\left(x_{\ell},t\right) & =\left(\dfrac{\alpha^{2}}{2\pi}\right)^{1/4}\dfrac{e^{ik_{0}\left(x_{\ell}-x_{s}\right)-iE_{k_{0}}t/\hbar}}{\sqrt{\alpha^{2}+itE_{k_{0}}^{''}/\left(2\hbar\right)}}\nonumber \\
 & \times\exp\left[-\frac{1}{4}\frac{\left(x_{\ell}-x_{s}-tE_{k_{0}}^{'}/\hbar\right)}{\alpha^{2}+itE_{k_{0}}^{''}/\left(2\hbar\right)}\right]\label{packet_bloch_time}
\end{align}
where $E_{k_{0}}^{'}$ and $E_{k_{0}}^{''}$ refer to the first and
second derivatives of the dispersion relation $E_{k}$ with respect
to wave vector $k$, evaluated at $k_{0}$. For nearest-neighbour
hopping the dispersion relation is given by $E_{k}=-2t_{0}\cos(ka)$.
It follows in this case that the spread of the wave packet, as a function
of time, will be constant for $k_{0}a=\pi/2$, and is given by the
initial width $\alpha$.

\begin{figure*}

\includegraphics[width=3in]{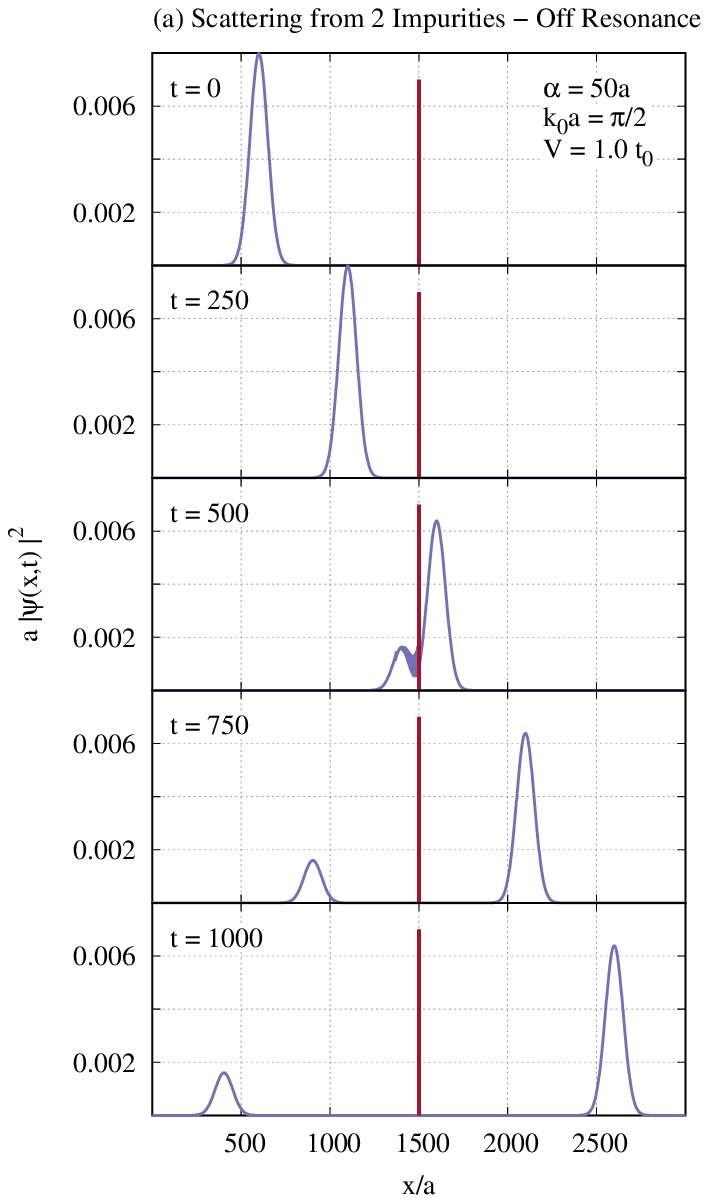} 
\includegraphics[width=3in]{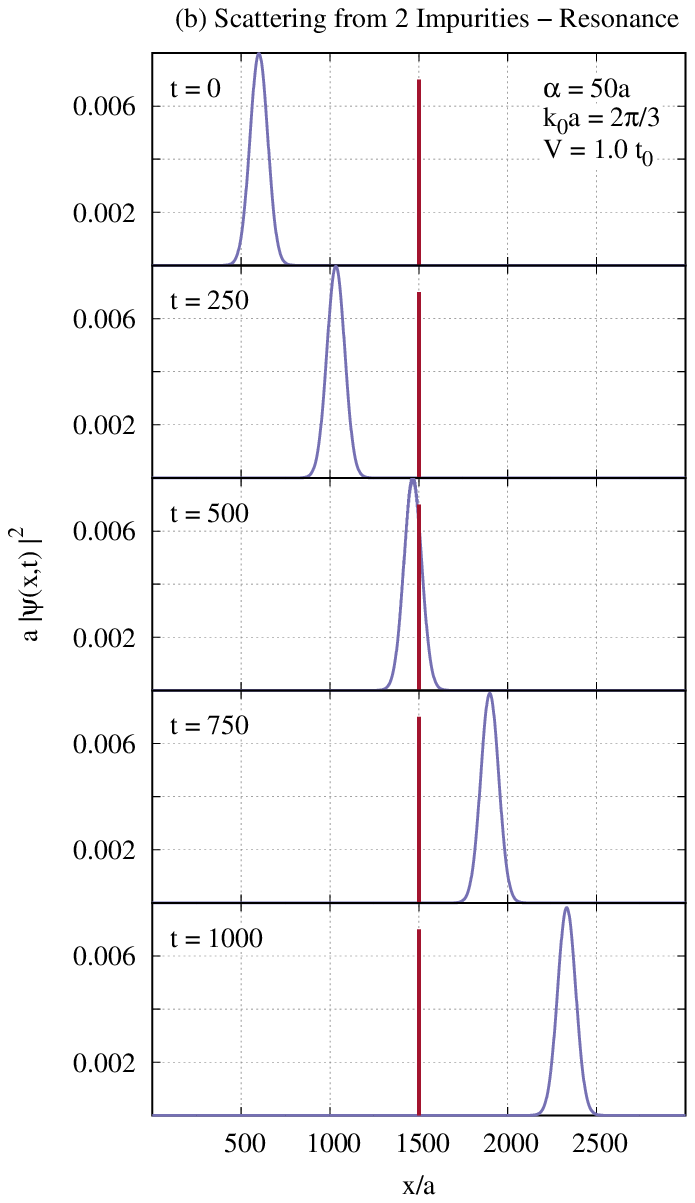}

\caption{
\label{Fig2 pi by 2 off resonance}
Plots of $a|\psi(x_{\ell},t)|^{2}$
as a function of lattice site for five different times (in units of
$\hbar/t_{0}$) for (a) off resonance and (b) on resonance. The barrier (vertical bars) consists
of two impurities, each with strength $V=1.0t_{0}$ located at sites
$1500$ and $1501$. We have illustrated wave packets for times well
before scattering (the first two) and well after scattering (the last
two). The wave-packet is initially
centered at $x_{\ell}=600a$. In plot (a)  $k_{0}a=\pi/2$ and no spreading
occurs because of the choice of $k_{0}$, but a significant amount
of reflection does occur. Note that no spreading
occurs after scattering, in both the transmitted and reflected wave
packets.  In plot (b)  $k_{0}a=2\pi/3$ and  is chosen
according to the resonance condition Eq.~(\ref{resonance condition})
so that no reflection should occur, as appears to be the case here. On the other hand, wave packet broadening occurs as a function of
time, as is (just barely) evident in the decreasing amplitude as a
function of time. The actual presence of a small reflected component
is shown in Fig.~\ref{Fig4 2pi by 3 zoom} below.
In both cases, the scattering is elastic and the reflected wave packet
propagates to the left and right with the same initial wave vector
magnitude, $k_{0}$.
}
\end{figure*}

As an example, in Fig.~\ref{Fig2 pi by 2 off resonance}(a) we show
a sequence of snapshots of a travelling wave packet, with $k_{0}a=\pi/2$,
that encounters two ($N=2$) barriers set side-by-side ($m=1$) at
$x_{\ell}=1500a$ and $x_{\ell}=1501a$. No spreading is predicted
to take place on an empty lattice. Two snapshots ($t=0$ and $t=250\hbar/t_{0}$)
are shown before scattering takes place and two snapshots ($t=750\hbar/t_{0}$
and $t=1000\hbar/t_{0}$) are shown for times after scattering. 
\bluee{``Before'' and ``after'' here are loosely defined and are meant to signify that the wave packet
profile is entirely to the left and right of the barrier region, respectively, as viewed by eye in such a figure.
Since a Gaussian has tails this is necessarily a loose definition.}  
The middle panel, showing the scattering event as the wave packet strikes
the barrier, displays considerable disruption in the wave packet.
More detailed descriptions of these transient phenomena can be found
in the simulations in Ref.~\onlinecite{staelens21}. It should be
clear from the two times chosen before and after the scattering, that
no spreading takes place, as the peak heights remain the same. In
particular it is true that no spreading takes place in either the
transmitted or reflected wave packets. A sizable reflectance takes
place in Fig.~\ref{Fig2 pi by 2 off resonance}(a) of approximately
20\%. From the curve shown in Fig.~\ref{Fig1} ($N=2$, $m=1$) it
is clear that the transmission is indeed expected to be approximately
80\% for this barrier configuration.

In contrast, in Fig.~\ref{Fig2 pi by 2 off resonance}(b) we show snapshots
of a wave packet with $k_{0}a=2\pi/3$, for which spreading is anticipated
based on Eq.~(\ref{packet_bloch_time}). Referring back to Fig.~\ref{Fig1}
($N=2$, $m=1$), we see that unit transmission is anticipated for
this wave vector. Sure enough, inspection of Fig.~\ref{Fig2 pi by 2 off resonance}(b)
shows a small decrease in amplitude of the wave packet due to spreading
as time progresses. Notice that the wave packet travels a little more
slowly than the one in Fig.~\ref{Fig2 pi by 2 off resonance}(a) since
the group velocity, $v_{{\rm gr}}(k)\equiv (dE_{k}/dk)/\hbar=2t_{0}a\sin(ka)/\hbar$,
is maximal at $k_{0}a=\pi/2$. It appears that on this scale, no
reflectance occurs, in apparent agreement with Fig.~\ref{Fig1} ($N=2$,
$m=1$).

\begin{figure}
\includegraphics[width=3in]{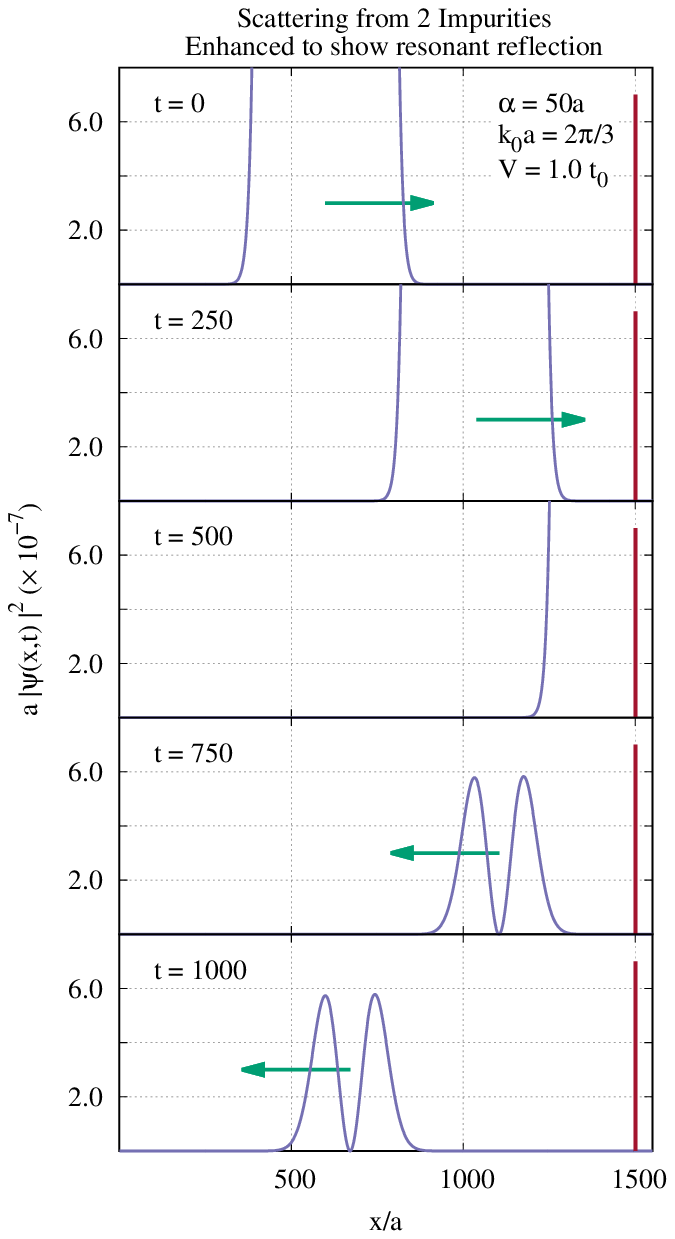}

\caption{\label{Fig4 2pi by 3 zoom}
An enlarged plot to show the reflected
component of $a\,|\psi(x_{\ell},t)|^{2}$ that is present in Fig.~\ref{Fig2 pi by 2 off resonance}(b). 
See Fig.~\ref{Fig2 pi by 2 off resonance}(b) for the initial wave packet and barrier parameters.
Note the vertical scale here, consistent with the fact that the reflected
profile in the two lowest panels were \textbf{not} visible in Fig.~\ref{Fig2 pi by 2 off resonance}(b).
Note also that the reflected wave packet profile ($t=750\hbar/t_{0}$
and $1000\hbar/t_{0}$) \textit{does not} consist of a single Gaussian
wave packet; rather, \textit{two} peaks are present for each time,
and the expected centroid of the reflecting wave packet is precisely
at the node between the two peaks. The reflected wave packet profile,
though non-Gaussian, remains constant as a function of time, as seen
visually by comparing the profile for the two times shown in the lowest
two panels. }
\end{figure}

A closer look, however, indicates that some reflectance does indeed
occur, as is illustrated in Fig.~\ref{Fig4 2pi by 3 zoom}, which
is  a greatly expanded version of Fig.~\ref{Fig2 pi by 2 off resonance}(b).
Several characteristics are of note. First, the reflectance is non-zero.
In fact, the amount of the wave packet reflected is indeed very tiny
(see the scale!) and can readily be adjusted by changing the width
$\alpha$ of the initial wave packet. These results are shown for
$\alpha=50a$, but a larger choice of the width $\alpha$ would 
reduce the reflectance accordingly. Essentially, the spread in wave vector about
$k_{0}a=2\pi/3$ is inversely proportional to $\alpha$. The larger
$\alpha$ is, the more the wave packet approaches a mono-energetic
wave packet with \textit{only} $k_{0}a=2\pi/3$ and therefore perfect
unit transmission (see Fig. 4 in Ref. \onlinecite{staelens21}). 

Secondly, the reflected component is no longer Gaussian; in fact it
appears to be a split wave packet with the peaks travelling coherently.
Thirdly, distorted as these split wave packets are from the
original in Fig.~\ref{Fig2 pi by 2 off resonance}(b), they nonetheless
retain their shape as a function of time. Note that \bluee{because $k_{0}a=2\pi/3$, the wave packets} do spread
slightly with time, though it is difficult to see on this figure. \bluee{For  $k_{0}a=\pi/2$ they do not spread at all,
and the split Gaussian remains as it is for all the times shown. It is this remarkable line-shape that we want
to more fully understand.} 
As we will show in the next section, this split wave function can
be described analytically by a single function that travels with wave vector
$k_{0}$.

\subsection{Analytical approach to describe reflected and transmitted wave packets}

The analytical expression for a wave packet in Eq.~(\ref{packet_bloch_time})
follows from an integration over all wave vectors centered around
$k\approx k_{0}$. The spread of the wave packet in real space $\alpha$
assures that only a small domain of wave vectors ($\Delta k\propto1/\alpha$)
surrounding $k_{0}$ is required, so that the energy dispersion can
also be expanded around $k_{0}$.

A similar expansion applies to both transmitted and reflected components
(see Eq.~(12) in Ref.~\onlinecite{kim0674}). The reflected component
is
\begin{equation}
\psi_{R}\left(x_{\ell},t\right)=\frac{1}{\sqrt{2\pi}}\int_{-\infty}^{+\infty}dk\,\phi_{-k_{0}}\left(k\right)\rho\left(k\right)e^{ikx}e^{-iE_{k}t/\hbar},\label{reflect_eq}
\end{equation}
where the limits of integration have been extended to $\pm\infty$,
$\rho(k)$ is the reflectance amplitude, and 
\begin{equation}
\phi_{-k_{0}}\left(k\right)=\left(\frac{2\alpha^{2}}{\pi}\right)^{1/4}\exp\left[-\alpha^{2}\left(k+k_{0}\right)^{2}\right].\label{phi_k_0}
\end{equation}
The reflectance amplitude can be determined from the plane wave theory
established in the previous Section \ref{sec:plane_wave}. Since this
is an elastic scattering event, the reflected wave packet travels
with the same energy as before the scattering event but in the negative
$x$ direction. The amplitude $\phi_{-k_{0}}(k)$ is hence indicated
to have been evaluated for $-k_{0}$, as will be the reflectance amplitude
$\rho(k)$ and the energy dispersion $E_{k}$.

By expanding $E(k)$ and $\rho(k)$ in Eq.~(\ref{reflect_eq}) around
$-k_{0}$, using the first order correction of $\rho(k)$, and the
Gaussian form of $\phi(k)$, we can explicitly calculate the reflected
probability $\left|\psi_{R}(x_{\ell},t)\right|^{2}$ to first order.
This approximation of Eq.~(\ref{reflect_eq}) yields 
\begin{align}
 & \psi_{R}\left(x_{\ell},t\right)\approx\frac{\rho'\left(-k_{0}\right)}{\sqrt{2\pi}}\left(\frac{2\alpha^{2}}{\pi}\right)^{1/4}e^{-iE_{0}t/\hbar}e^{-ixk_{0}}\nonumber \\
 & \times\int_{-\infty}^{+\infty}dk(k+k_{0})e^{-\left[\alpha^{2}+iE''_{0}t/2\hbar\right]\left(k+k_{0}\right)^{2}}e^{i\left(x-E'_{0}t/\hbar\right)(k+k_{0})}.
\label{reflect_eq-approx}
\end{align}
The above integral is Gaussian and can be done analytically. The probability
is given by 
\begin{eqnarray}
\left|\psi_{R}\left(x_{\ell},t\right)\right|^{2} & \approx & \frac{\left|\rho'\left(k_{0}\right)\right|^{2}}{\sqrt{2\pi}}\frac{y^{2}\left(t\right)e^{-y^{2}\left(t\right)}}{2\alpha^{2}\sqrt{\alpha^{2}\left[1+u^{2}\left(t\right)\right]}},
\label{psi^2 analytic}
\end{eqnarray}
where 
\begin{equation}
u\left(t\right)\equiv\frac{E_{k_{0}}^{''}t}{2\hbar\alpha^{2}},
\end{equation}
and
\begin{equation}
y\left(t\right)\equiv\frac{\left(x_{\ell}-E_{k_{0}}^{'}t/\hbar\right)}{\sqrt{2\alpha^{2}\left[1+u^{2}\left(t\right)\right]}}.
\end{equation}
Here, $E^\prime_{k_0} \equiv dE_k/dk$ evaluated at $k = k_0$, and similarly for $E^{''}_{k_0}$.

Equation (\ref{psi^2 analytic}) describes a ``split'' Gaussian
function. The Gaussian is split because of the factor of $y^{2}(t)$
that precedes the exponential. The lower two panels in Fig.~\ref{Fig4 2pi by 3 zoom}
make clear what a split Gaussian looks like. Note that for $k_{0}a=\pm\pi/2$,
$u(t)=0$ since $E_{k_{0}}^{''}=0$, so this expression describes
a travelling split Gaussian whose width remains constant as a function
of time for these particular wave vectors.

\begin{figure}
\includegraphics{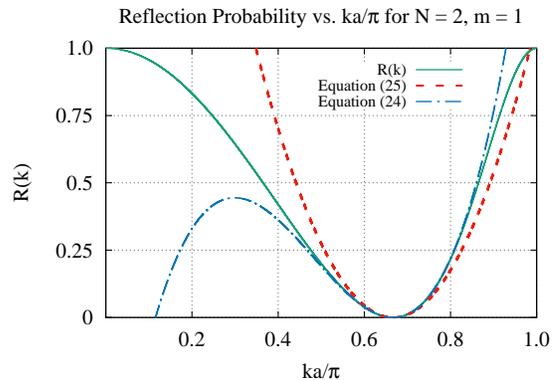}

\caption{\label{fig5}Plot of the reflection probability $R(k)$ vs. $ka/\pi$
(solid green curve). Also shown for comparison is the first order
correction (Eq.~(\ref{reflection_probability}), dotted red curve)
along with the next order correction (Eq.~(\ref{first_order_correction}),
dot-dashed blue curve). The range of agreement is clearly extended
by the next order correction. The parameters here are $k_{0}a=2\pi/3$,
$V=t_{0}$, $N=2$, and $m=1$.}
\end{figure}

To understand how well Eq.~(\ref{psi^2 analytic}) works we examine
the reflection probability $R(k)$, which is determined from the reflectance
amplitude as $R(k)=\left|\rho(k)\right|^{2}$. The reflectance amplitude
is given by $\rho(k)=M_{12}/M_{22}$ as defined in  the online Supplementary Material~\ref{A:Transfer-Matrix}.
For the specific case $N=2$ and $m=1$, we can use the derived elements
of the transfer matrix in Eq.~(\ref{eq:dimer M elements}) to get
\begin{equation}
\rho(k)= i e^{2ika}{f(k) \over 1 - if(k)}
\label{rho_n2_m1}
\end{equation}
from which Eq.~(\ref{reflection}) immediately follows. The specific form of $f(k)$ for $N=2$ and $m=1$ is given below.
Since we choose $k_{0}$ so that $\rho(k_{0})=0$, Eq.~(\ref{reflect_eq-approx})
came from the first order expansion, $\rho(k_{0}) \approx \rho^{\prime}(k_{0})(k-k_{0})$,
and therefore 
\begin{equation}
R\left(k\right)\approx\left|\rho'\left(k_{0}\right)\right|^{2}\left(k-k_{0}\right)^{2}.\label{reflection_probability}
\end{equation}
This result is shown for $k_{0}a=2\pi/3$ in Fig.~(\ref{fig5}) (dashed red curve)
since this is the wave vector for resonance to occur with $V=t_{0}$.
This approximate expression for the reflection probability clearly
compares favorably with the full reflection probability $R(k)$ computed
from Eq.~(\ref{reflection}) (solid green curve) \bluee{only} for a limited range
of wave vectors close to $k_{0}$. 

Going to higher order in $k-k_0$  is possible and straightforward. For example,
using that $f(k)\equiv v_{k}U_{N-1}(h_{k})$, for the case $N=2$
and $m=1$ we find
\begin{equation}
f(k)=\frac{V\left(\cos\left(ka\right)+V/2t_{0}\right)}{t_{0}\sin\left(ka\right)}.
\end{equation}
By expanding both $\cos(ka)$ and $\sin(ka)$ around $k_{0}$ in the
above expression, and by using the resonant condition Eq.~(\ref{resonance condition})
that gives us $\cos(k_{0}a)+V/2t_{0}=0$, we obtain
\begin{eqnarray}
f(k) & \approx & \frac{V/t_{0}}{\sin\left(k_{0}a\right)+a\left(k-k_{0}\right)\cos\left(k_{0}a\right)}\nonumber \\
 & \times & \left[\cos\left(k_{0}a\right)-a\left(k-k_{0}\right)\sin\left(k_{0}a\right)\right.\nonumber \\
 &  & \left.-\frac{1}{2}a^2\left(k-k_{0}\right){}^{2}\cos\left(k_{0}a\right)+\frac{V}{2t_{0}}\right]\nonumber \\
 & \approx & -\frac{V}{t_{0}}a\left(k-k_{0}\right)\left[1-\frac{1}{2}a\left(k-k_{0}\right)\frac{\cos\left(k_{0}a\right)}{\sin\left(k_{0}a\right)}\right],\label{fofk}
\end{eqnarray}
where the correction in the next order of $a(k-k_{0})$ is now included.
Substituting Eq.~(\ref{fofk}) into Eq.~(\ref{reflection}), expanding about $k_{0}$
and evaluating the coefficients at $k_{0}a=2\pi/3$ and $V=t_{0}$
results in 
\begin{equation}
R(k)\approx a^2\left(k-k_{0}\right)^{2}\left[1+\frac{1}{\sqrt{3}}a\left(k-k_{0}\right)\right],\label{first_order_correction}
\end{equation}
which is also shown in Fig.~\ref{fig5} as the dot-dashed blue
curve. This results agrees with the full plane-wave result from Eq.~(\ref{transmission})
over a more extend range of $k$ values around $k_{0}$ compared to
the first order result. This correction may be useful as the wave
packet width decreases, requiring more and more accuracy in the reflection
coefficient for wave vectors further away from $k_{0}$. Nonetheless,
in the comparisons that follow we will use Eq.~(\ref{reflection_probability}),
which corresponds to just the first term in Eq.~(\ref{first_order_correction}).

\subsection{More detailed comparisons}

\begin{figure*}

\includegraphics{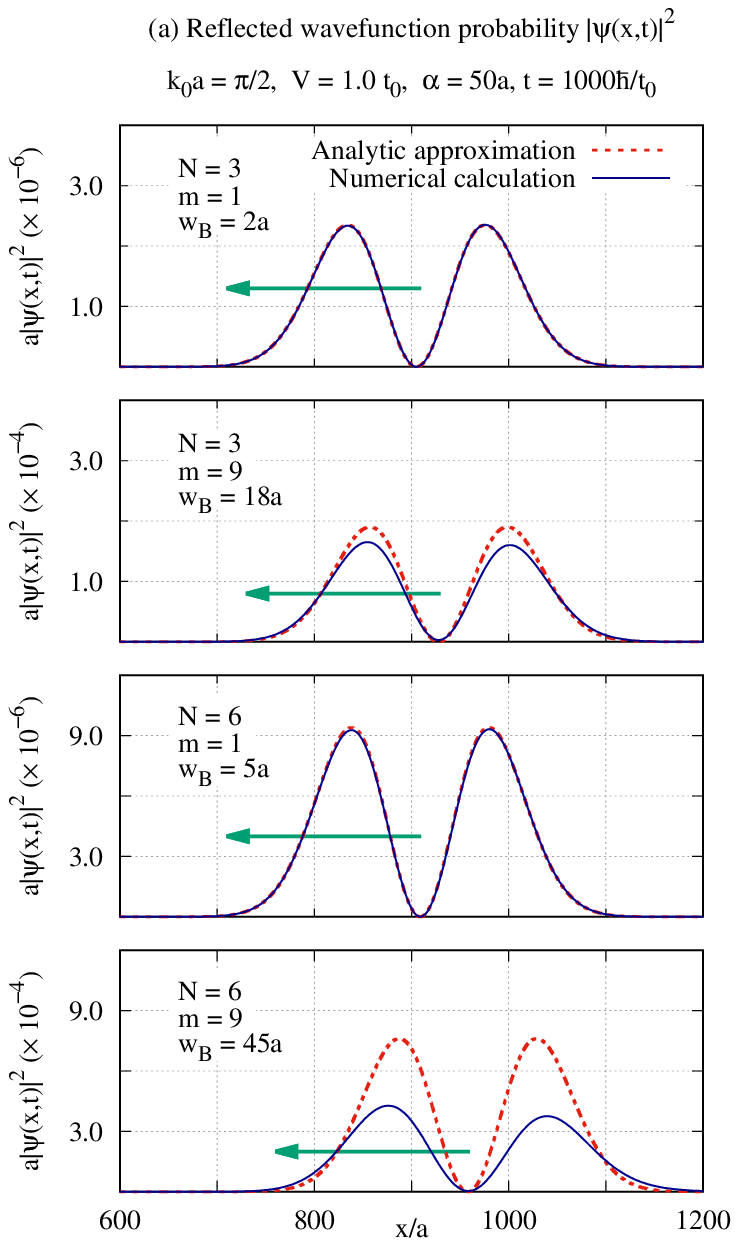}
\includegraphics{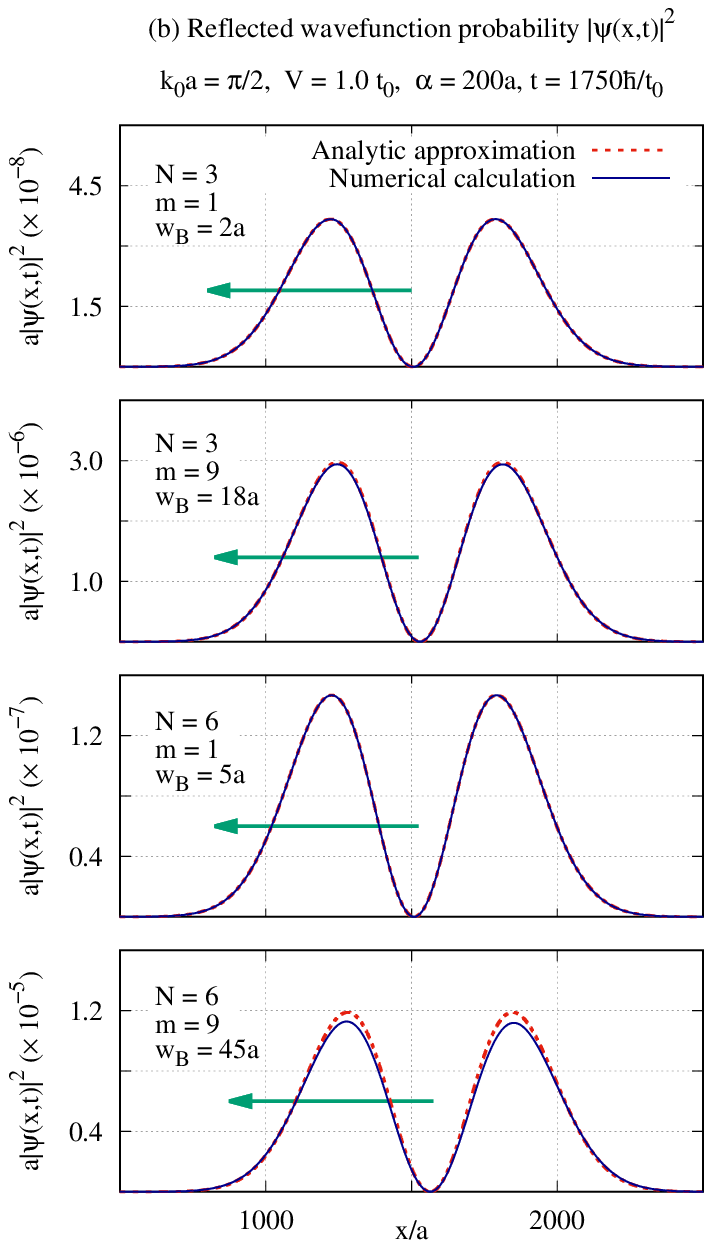}

\caption{\label{fig6}
The reflected component of $a|\psi(x_{\ell},t)|^{2}$,  after scattering has taken place at the barrier,
as a function of lattice site for $t=1000\hbar/t_{0}$ and the various
barrier configurations as shown. The barrier starts at $x=1500a$
(off-scale to the right). The results are determined both numerically
(solid blue curves) and analytically through Eq.~(\ref{psi^2 analytic})
(dashed red curves). The wave packet has $k_{0}a=\pi/2$,
and starts at $t=0$ from $x_{0}=600a$, with (a)  $\alpha=50a$ and (b)  $\alpha=200a$. 
 In
all these cases the transmission resonance condition is satisfied
(see Eq.~(\ref{fofk2})). Note
the vertical scale in this figure, consistent with the expectation that these components are nominal.
All are in qualitative agreement and quantitatively improve as the wave packet width is increased with respect to the barrier length.
}
\end{figure*}

In general, as the barrier width increases (i.e. as both $N$ and
$m$ increase), we expect that the accuracy of this approximation
will deteriorate (see Fig.~\ref{Fig1}). We wish to assess the analytic
expression for the reflection probability given by Eq.~(\ref{psi^2 analytic}),
for a variety of barriers. For this purpose, we show in Fig.~\ref{fig6}
the reflection probability as a function of position at some time
(here, $t=1000\hbar/t_{0}$) \textit{well after} the incoming wave
packet has scattered off the barrier. Note the vertical scale; we are zooming in on a very tiny amount of
reflected wave packet. As in Fig.~\ref{Fig4 2pi by 3 zoom} we are
using resonant scattering conditions in all cases, (i) $N=3$, $m=1$,
(ii) $N=3$, $m=9$, (iii) $N=6$, $m=1$, and (iv) $N=6$, $m=9$,
but here $k_{0}a=\pi/2$ and $V=1.0t_{0}$ for all barriers. The solid
blue curve denotes the numerical calculation, while the dashed red
curve denotes the analytical approximation given by Eq.~(\ref{psi^2 analytic}).
As is evident, when the width of the wave packet $\alpha$ greatly
exceeds the width of the barrier region $w_{B}$ (cases (i) and (iii)),
the approximation is excellent. As the barrier region width increases
(cases (ii) and (iv)), the approximation begins to fail. Note that
we are using $k_{0}a=\pi/2$, so there is no deterioration of the
Gaussian (or split Gaussian in the case of the reflected amplitude)
as a function of time, and an identical result would hold for a later
time as well. \bluee{The accuracy of the analytical approximation in the two cases where the extension of the
barrier is not so large (first and third panels) is remarkable given that the accuracy of the reflection probability
used is similar to that shown in Fig.~\ref{fig5}, and clearly works on a rather limited range about $k_0$.}

\begin{figure}
\includegraphics{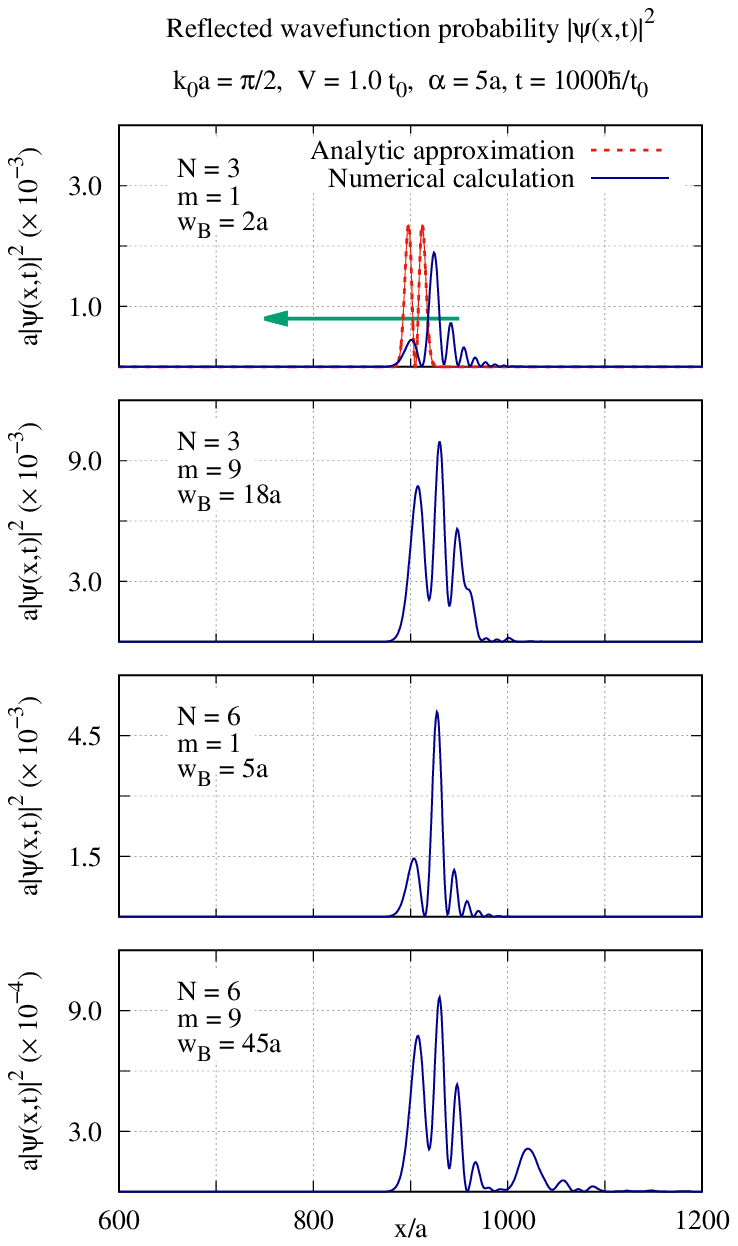}

\caption{\label{fig8} As Fig.~\ref{fig6} but
now for a narrower wave packet with $\alpha=5a$. As before
we show only the reflected components of $a|\psi(x_{\ell},t)|^{2}$
as a function of lattice site for $t=1000\hbar/t_{0}$ and the various
barrier configurations as shown. The barrier starts at $x=1500a$
(off-scale to the right). The results are determined both numerically
(solid blue curves) and analytically through Eq.~(\ref{psi^2 analytic})
(dashed red curve in the first panel). The reflected portion is still
small, though the vertical scale is now several orders of magnitude
higher than in Fig.~\ref{fig6}. The numerical results have significantly
distorted shapes, and can no longer be described by split Gaussian
wave packets.}
\end{figure}

More general conditions for resonance can be readily determined. For
example, for $V=t_{0}$, there is a resonant transmission that occurs
at $k_{0}a=\pi/2$ in all cases in which $m$ is odd. A straightforward
expansion of $f(k)$ {[}defined following Eq.~(\ref{reflection}){]}
for $m$ odd, $V=t_{0}$ and $k_{0}a=\pi/2$ results in 
\begin{eqnarray}
f(k)\approx-2ma(k-k_{0}),\quad\textrm{for}\ N=3\\
f(k)\approx-4ma(k-k_{0}),\quad\textrm{for}\ N=6\label{fofk2}
\end{eqnarray}
and similar formulas can be derived for other cases.

As expected, the agreement indeed deteriorates both as $N$ increases
and as $m$ increases, as is apparent in Fig.~\ref{fig6}(a). At the
same time, we should remind the reader that we are capturing a tiny
reflected component, so in some sense the second and fourth panels
still show excellent qualitative agreement. For example, the fact
that the original simple Gaussian wave packet shape has now been divided
into a split Gaussian is captured magnificently by the approximate
result shown here, based on Eq.~(\ref{psi^2 analytic}). 

To illustrate how improved agreement can be obtained systematically
with increasing wave packet width, we show in Fig.~\ref{fig6}(b) the
results from Fig.~\ref{fig6}(a), but now for the wider wave packet
with $\alpha=200a$. There is clear quantitative improvement for all
panels, most evident in the second and fourth panels. Clearly one
of the issues with the poorer agreement in Fig.~\ref{fig6}(a) is that
the barrier width is comparable to the width of the wave packet. For example  in the fourth panel the barrier width
$w_{B}$ is
$45a$, which is very close to the width of the wave packet  $\alpha=50a$ that
is used. With a wave packet width of $\alpha=200a$, we arrive at
a situation where the wave packet width is much greater than any length
scale associated with the barrier, and of course, for still larger
values of $\alpha$ the agreement with the numerical results improves
still further.

Conversely, for smaller values of $\alpha$, the agreement deteriorates
considerably. In Fig.~\ref{fig8} we show the same four cases as
in Fig.~\ref{fig6}(a) and (b), but now with $\alpha=5a$.
Even for just three barriers (located at sites $1500$, $1501$, and
$1502$), there is no remnant of a split Gaussian, and in fact there
is even a deterioration of the lineshape as a function of time, indicating
that the expansion behind the analytical derivation of Eq.~(\ref{packet_bloch_time})
has lost its validity. The approximate result provided by Eq.~(\ref{psi^2 analytic})
is shown in the first panel only (dashed red curve), and is qualitatively
incorrect. The results for such a narrow wave packet cannot be described
with the help of the plane wave theory, as scattering events at the
individual components of the barriers are disconnected, and can no
longer interfere with one another.

\section{Summary}

We have formulated the one-dimensional scattering problem on a lattice
to study the effect of a non-infinite wave packet width on the transmission
and reflection coefficients. As much as possible, we have focussed
on the wave vector $k_{0}a=\pi/2$ so that the Gaussian wave packet
that we use retains the same width for extended periods of time.
We emphasize that having a constant width as a function of time is
an approximate result based on a Taylor expansion in the exponential,
but the numerical results presented in this paper do not use this
approximation, and nevertheless demonstrate this characteristic for
the time scales involved in the scattering process. This makes the
use of these wave packets a helpful tool, since all changes can be
attributed to the scattering process, and \textit{not} to the general
time evolution of the wave packet itself. \bluee{Moreover, conceptually, it
is easier to think of these wave packets as actual particles used in the beam experiment --- they
actually move towards and scatter off and move away from the target as a function of time. This is in contrast to the plane
wave description, where time has disappeared from the problem, and incoming, scattering and outgoing particles appear to 
the novice to be present at all times and in all places in a sort of steady-state condition.}

We also focussed on barrier potentials that result in resonant transmission
for a mono-energetic plane wave, namely $V=t_{0}$ for the four cases considered
in the last three figures. Then any reflection can be attributed to the
fact that a finite width wave packet was used in the scattering process.
We found numerically that a minute amount was reflected, but the reflected
component had a \bluee{well defined and} peculiar split Gaussian profile. Based on these results
we devised an approximation for the reflected component, Eq.~(\ref{psi^2 analytic})
which, when combined with a simple approximation for the reflectivity
as a function of wave vector, yields excellent results compared to
the exact numerical results. For reasonably large wave packet widths,
we find quantitative agreement for the reflected wave profile, and
a good understanding of its accuracy based on the initial Gaussian
width. In this way a combination of plane wave-based reflection coefficients
(Fig.~\ref{Fig1}), together with analytical wave packet dynamics
allows us a more insightful understanding of the \bluee{time dependence of the} scattering process
off of simple barriers. \bluee{At the same time, for wave packets whose width is less than the characteristic
barrier length scale, the scattering description in terms of plane waves breaks down completely, as the
results of Fig.~\ref{fig8} show.}

\section*{ACKNOWLEDGMENTS}

This work was supported in part by the Natural Sciences and Engineering
Research Council of Canada (NSERC) and by an MIF from the Province
of Alberta.\\

\noindent The authors have no conflicts to disclose.

\vfil\eject

\noindent \phantom{aaaaaaaaaaa} {\bf Supplementary Material}

\appendix

\section{Transfer Matrix\label{A:Transfer-Matrix}}

The solution to the one-dimensional quantum mechanical scattering
problem in the presence of a localized potential $V\left(x\right)$
using the transfer matrix method has been investigated extensively.\citep{phillips03,markos08,joannopoulos08,griffiths92,sprung93,griffiths01}
We review this work in this Appendix, both to keep this work self-contained,
and to define various terms for the specific application of wave packet
propagation in the tight binding model.

The transfer matrix relates the amplitudes of propagating waves between
the left and right side of a barrier $V$. In the continuum, a general
solution to the time-independent Schr\"odinger equation 
\begin{equation}
-\frac{\hbar^{2}}{2m}\frac{d^{2}\psi(x)}{dx^{2}}+V\left(x\right)\psi\left(x\right)=E\psi\left(x\right),
\end{equation}
is required, where $V(x)$ is defined on the interval $(a,b)$. In
general the wave function is given by 
\begin{equation}
\psi\left(x\right)=\begin{cases}
Ae^{ikx}+Be^{-ikx} & {\rm for}\ \ x<a\\
\psi_{ab}\left(x\right) & {\rm for}\ \ a<x<b\\
Ce^{ikx}+De^{-ikx} & {\rm for}\ \ x>b.
\end{cases}
\end{equation}
The scattering matrix $\boldsymbol{S}$ relates the incoming and outgoing
wave function components as 
\[
\left(\begin{array}{c}
B\\
C
\end{array}\right)=\boldsymbol{S}\left(\begin{array}{c}
A\\
D
\end{array}\right)=\left(\begin{array}{cc}
S_{11} & S_{12}\\
S_{21} & S_{22}
\end{array}\right)\left(\begin{array}{c}
A\\
D
\end{array}\right),
\]
while the transfer matrix $\boldsymbol{M}$ relates the wave function
components between the left side and right side of the barrier as
\[
\left(\begin{array}{c}
C\\
D
\end{array}\right)=\boldsymbol{M}\left(\begin{array}{c}
A\\
B
\end{array}\right)=\left(\begin{array}{cc}
M_{11} & M_{12}\\
M_{21} & M_{22}
\end{array}\right)\left(\begin{array}{c}
A\\
B
\end{array}\right).
\]
These two descriptions above are related as\citep{markos08}
\begin{equation}
\boldsymbol{M}=\begin{pmatrix}S_{21}-\frac{S_{22}S_{11}}{S_{12}} & \frac{S_{22}}{S_{12}}\\
-\frac{S_{11}}{S_{12}} & \frac{1}{S_{12}}
\end{pmatrix}.\label{eq:matrixM}
\end{equation}
The elements of the scattering matrix $\boldsymbol{S}$ give us the
transmission amplitude from right to left of the barrier $\tau=S_{12}$
and left to right $\tau'=S_{21}$, and the reflection amplitudes on
each side of the barrier $\rho=S_{22}$ and $\rho'=S_{11}$. We
can write the scattering matrix as 
\begin{equation}
\boldsymbol{S}=\left(\begin{array}{cc}
\rho' & \tau\\
\tau' & \rho
\end{array}\right).
\end{equation}
The transmission (reflection) coefficient is then defined as the probability
that a particle is transmitted (reflected) by $T=\left|\tau\right|^{2}$
($R=\left|\rho\right|^{2}$).

Furthermore, it can be shown\citep{markos08} that time
reversal symmetry and conservation of current density further restrict
the elements of $\boldsymbol{M}$ and $\boldsymbol{S}$. Following
Ref.~\onlinecite{markos08}, time-reversal symmetry leads to the
conclusion that the scattering matrix $\boldsymbol{S}$ is unitary.
This gives us the relation (one of four, see Eq.~(1.26) in Ref.~\onlinecite{markos08})
that $S_{11}^{*}S_{12}+S_{12}^{*}S_{22}=(\rho'){}^{*}\tau + (\tau'){}^{*}\rho=0$.
This lets us write 
\begin{equation}
\frac{\rho}{\tau}=-\left(\frac{\rho'}{\tau'}\right)^{*},
\end{equation}
 and we can use this to simplify the element $M_{11}$ in Eq.~(\ref{eq:matrixM})
as 
\begin{equation}
M_{11}=\tau'-\frac{\rho\rho'}{\tau}=\tau'+\left(\frac{\rho'}{\tau'}\right)^{*}\rho'=\frac{\left|\tau'\right|^{2}+\left|\rho'\right|^{2}}{(\tau')^{*}}.
\end{equation}
Conservation of current density leads to the relation that $\left|\tau'\right|^{2}+\left|\rho'\right|^{2}=\left|\tau\right|^{2}+\left|\rho\right|^{2}=1$,
which lets us simplify the transfer matrix to the following form

\begin{equation}
\boldsymbol{M}=\begin{pmatrix}\frac{1}{(\tau')^{*}} & \frac{\rho}{\tau}\\
-\frac{\rho'}{\tau} & \frac{1}{\tau}
\end{pmatrix}.\label{eq: simplified transfer matrix M}
\end{equation}

\section{Tight-Binding Model and Single Impurity\label{A:Tight-Binding-Model}}

The tight-binding Hamiltonian for a 1D lattice with $N$ impurities
is 
\begin{eqnarray}
H & = & -t_{0}\sum_{\ell}\left[c_{\ell}^{\dagger}c_{\ell+1}+c_{\ell+1}^{\dagger}c_{\ell}\right]+\sum_{\ell=1}^{N}V_{\ell}c_{\ell}^{\dagger}c_{\ell}
\end{eqnarray}
and the corresponding Schr\"odinger equation is 
\begin{eqnarray}
-t_0\left[\psi_{\ell+1}+\psi_{\ell-1}\right]+V_{\ell}\psi_{\ell} & = & E_{k}\psi_{\ell}.\label{eq:Schrodinger tight binding}
\end{eqnarray}
This Hamiltonian assumes nearest-neighbor hopping only. Equivalently,
the Schr\"odinger equation (Eq.~\ref{eq:Schrodinger tight binding})
can be written as the matrix equation 
\begin{eqnarray}
\begin{pmatrix}\psi_{\ell+1}\\
\psi_{\ell}
\end{pmatrix} & = & \begin{pmatrix}\tilde{V}_{\ell}-\tilde{E}_{k} & -1\\
1 & 0
\end{pmatrix}\begin{pmatrix}\psi_{\ell}\\
\psi_{\ell-1}
\end{pmatrix}\label{eq:Schr. matrix equation}\\
 & = & \boldsymbol{P}_{\ell}\begin{pmatrix}\psi_{\ell}\\
\psi_{\ell-1}
\end{pmatrix}.
\end{eqnarray}
where $\tilde{V}\equiv V/t_{0}$ and $\tilde{E}=E/t_{0}$. We assume
that a wave function at a site $\ell$ (at coordinate $x_{\ell}$)
is of the form $\psi_{\ell}=Ae^{ikx_{\ell}}+Be^{-ikx_{\ell}}$.

Our aim is to find the transfer matrix that relates the amplitudes
of a wave function that satisfies Eq.(\ref{eq:Schrodinger tight binding})
at lattice site $x_{\ell}$ with those at the nearest neighbour site
$x_{\ell+1}$. So if $\psi_{\ell}=Ae^{ikx_{\ell}}+Be^{-ikx_{\ell}}$
and $\psi_{\ell+1}=Ce^{ikx_{\ell+1}}+De^{-ikx_{\ell+1}}$, we define
a matrix $\boldsymbol{M}_{\ell}$ that satisfies 
\begin{eqnarray}
\begin{pmatrix}Ce^{ikx_{\ell+1}}\\
De^{-ikx_{\ell+1}}
\end{pmatrix} & = & \boldsymbol{M}_{\ell}\begin{pmatrix}Ae^{ikx_{\ell}}\\
Be^{-ikx_{\ell}}
\end{pmatrix},\label{eq:trasfer equation}
\end{eqnarray}
We now write $\boldsymbol{M}_{n}$ in terms of the Schr\"odinger equation
Eq.~(\ref{eq:Schrodinger tight binding}).

The wave function at lattice site $\ell$, $\psi_{\ell}=Ae^{ikx_{\ell}}+Be^{-ikx_{\ell}}=Ae^{ik\ell a}+Be^{-ik\ell a}$,
can be expressed as the matrix equation 
\begin{eqnarray}
\begin{pmatrix}\psi_{\ell}\\
\psi_{\ell-1}
\end{pmatrix} & = & \begin{pmatrix}1 & 1\\
e^{-ika} & e^{ika}
\end{pmatrix}\begin{pmatrix}Ae^{ik\ell a}\\
Be^{-ik\ell a}
\end{pmatrix}\label{eq: wavefunction}\\
 & = & \boldsymbol{Q}\begin{pmatrix}Ae^{ik\ell a}\\
Be^{-ik\ell a}
\end{pmatrix},
\end{eqnarray}
where $\boldsymbol{Q}$ is just the matrix to give back the correct
form of $\psi_{\ell}$ and $\psi_{\ell-1}$, and similarly for $\psi_{\ell+1}$,
\begin{eqnarray}
\begin{pmatrix}\psi_{\ell+1}\\
\psi_{\ell}
\end{pmatrix} & = & \begin{pmatrix}1 & 1\\
e^{-ika} & e^{ika}
\end{pmatrix}\begin{pmatrix}Ce^{ik\left(\ell+1\right)a}\\
De^{-ik\left(\ell+1\right)a}
\end{pmatrix}\\
 & = & \boldsymbol{Q}\begin{pmatrix}Ce^{ik\left(\ell+1\right)a}\\
De^{-ik\left(\ell+1\right)a}
\end{pmatrix}.
\end{eqnarray}
We can use the above expressions and Eq.~(\ref{eq:Schr. matrix equation})
to rewrite Eq.~(\ref{eq:trasfer equation}) as 
\begin{eqnarray}
\begin{pmatrix}Ce^{ik\left(\ell+1\right)a}\\
De^{-ik\left(\ell+1\right)a}
\end{pmatrix} & = & \boldsymbol{Q}^{-1}\begin{pmatrix}\psi_{\ell+1}\\
\psi_{\ell}
\end{pmatrix}\\
 & = & \boldsymbol{Q}^{-1}\boldsymbol{P}_{\ell}\begin{pmatrix}\psi_{\ell}\\
\psi_{\ell-1}
\end{pmatrix}\\
 & = & \boldsymbol{Q}^{-1}\boldsymbol{P}_{\ell}\boldsymbol{Q}\begin{pmatrix}Ae^{ik\ell a}\\
Be^{-ik\ell a}
\end{pmatrix}\\
 & = & \boldsymbol{M}_{\ell}\begin{pmatrix}Ae^{ik\ell a}\\
Be^{-ik\ell a}
\end{pmatrix}
\end{eqnarray}
where the transfer matrix is given by $\boldsymbol{M}_{\ell}=\boldsymbol{Q}^{-1}\boldsymbol{P}_{\ell}\boldsymbol{Q}$.

Using the explicit forms of $\boldsymbol{P}_{\ell}$ and $\boldsymbol{Q}$
we calculate $\boldsymbol{M}_{\ell}$ explicitly 
\begin{widetext}
\begin{eqnarray}
\boldsymbol{M}_{\ell} & = & \boldsymbol{Q}^{-1}\boldsymbol{P}_{\ell}\boldsymbol{Q}\\
 & = & \frac{1}{2i\sin\left(ka\right)}\begin{pmatrix}e^{ika} & -1\\
-e^{-ika} & 1
\end{pmatrix}\begin{pmatrix}\tilde{V}_{\ell}-\tilde{E}_{k} & -1\\
1 & 0
\end{pmatrix}\begin{pmatrix}1 & 1\\
e^{-ika} & e^{ika}
\end{pmatrix}\nonumber \\
 & = & \frac{1}{2i\sin\left(ka\right)}\begin{pmatrix}-2+e^{ika}\left(\tilde{V}_{\ell}-\tilde{E}_{k}\right) & -1-e^{2ika}+e^{ika}\left(\tilde{V}_{\ell}-\tilde{E}_{k}\right)\\
1+e^{2ika}-e^{-ika}\left(\tilde{V}_{\ell}-\tilde{E}_{k}\right) & 2-e^{-ika}\left(\tilde{V}_{\ell}-\tilde{E}_{k}\right)
\end{pmatrix}
\end{eqnarray}
\end{widetext}

In the tight-binding model, the dispersion relation is given by $E_{k}=-2t_{0}\cos(ka)$.
Substituting this into the above equation (with $t_{0}=1$) and simplifying
gives us the transfer matrix for lattice site $\ell$, 
\begin{align}
 & \boldsymbol{M}_{\ell}\nonumber \\
 & =\begin{pmatrix}e^{ika}\left[1+\frac{\tilde{V}_{\ell}}{2i\sin\left(ka\right)}\right] & e^{ika}\frac{\tilde{V}_{\ell}}{2i\sin\left(ka\right)}\\
-e^{-ika}\frac{\tilde{V}_{\ell}}{2i\sin\left(ka\right)} & e^{-ika}\left[1-\frac{\tilde{V}_{\ell}}{2i\sin\left(ka\right)}\right]
\end{pmatrix}.\label{eq: M expression}
\end{align}
The probability for transmission for a single impurity is then easily
calculated from the above expression as 
\begin{eqnarray}
T & = & =\frac{1}{\left|M_{22}\right|^{2}}=\frac{1}{1+\tilde{V}_{\ell}^{2}\frac{1}{4\sin^{2}\left(ka\right)}}\nonumber \\
 & = & \frac{\sin^{2}\left(ka\right)}{\sin^{2}\left(ka\right)+V_{\ell}^{2}/4t_{0}^{2}}.
\end{eqnarray}

For the case where there are no impurities the transfer matrix Eq.~(\ref{eq: M expression})
simplifies to 
\begin{equation}
\boldsymbol{M}^{\left(1\right)}=\begin{pmatrix}e^{ika} & 0\\
0 & e^{-ika}
\end{pmatrix}
\end{equation}
and gives a simple one lattice ``hop'' satisfying Eq.~(\ref{eq:trasfer equation}).
For $m$ impurity free hops this simply extends to multiplying $M^{(1)}$
$m$ times, which gives 
\begin{equation}
\boldsymbol{M}^{\left(m\right)}=\begin{pmatrix}e^{ika} & 0\\
0 & e^{-ika}
\end{pmatrix}^{m}=\begin{pmatrix}e^{ikma} & 0\\
0 & e^{-ikma}
\end{pmatrix}.\label{eq:free propagation}
\end{equation}

\section{Dimers and Generalization to $N$ Impurities\label{A:Dimers-and-Generalization}}

With the general transfer matrix for a single site $\ell$, we can
calculate the transfer matrix (and therefore the transmission and
reflection amplitudes) of a system of two impurities separated by
$m$ lattice sites. This configuration is referred to as a dimer and
has been explored extensively in Refs.~\onlinecite{kim0673,kim0674,dunlap90,wu91,wu92,datta93,giri93}.

Let $V_{\ell}$ be the potential at a site $\ell$, and $V_{\ell+m}$
the potential at a site $m$ lattice spacings away at $\ell+m.$ We
will assume that these two potentials are of equal strength $V_{\ell}=V_{\ell+m}=V$.
Between these two potentials there are $m-1$ sites where $V_{i}=0$.

For such a configuration the wave function amplitude on the left-hand-side
of the entire barrier (dimer) can be seen as encountering the first
impurity $V_{\ell}$, described by a transfer matrix $\boldsymbol{M}_{1}$,
freely propagating over $m-1$ lattice sites, described by $\boldsymbol{M}^{(m-1)}$,
and finally encountering the second impurity $V_{\ell+m},$ described
by $\boldsymbol{M}_{2}$. The transfer matrix describing the dimer
is given by $\boldsymbol{M}=\boldsymbol{M}_{1}\boldsymbol{M}^{(m-1)}\boldsymbol{M}_{2}$.

Since $V_{\ell}=V_{\ell+m}=V,$ we have $\boldsymbol{M}_{1}=\boldsymbol{M}_{2}$
given by Eq. (\ref{eq: M expression}) ($\tilde{V}=V/t_{0}$) 
\begin{align}
\boldsymbol{M}_{1} & =\boldsymbol{M}_{2}\nonumber \\
 & =\begin{pmatrix}e^{ika}\left(1+\frac{\tilde{V}}{2i\sin\left(ka\right)}\right) & e^{ika}\frac{\tilde{V}}{2i\sin\left(ka\right)}\\
-e^{-ika}\frac{\tilde{V}}{2i\sin\left(ka\right)} & e^{-ika}\left(1-\frac{\tilde{V}}{2i\sin\left(ka\right)}\right)
\end{pmatrix}.
\end{align}

\begin{widetext}
Calculating the transfer matrix explicitly for the system gives 
\begin{align}
\boldsymbol{M} & =\boldsymbol{M}_{1}\boldsymbol{M}^{\left(m-1\right)}\boldsymbol{M}_{2}\nonumber \\
 & =\begin{pmatrix}e^{ika}\left(1+\frac{\tilde{V}}{2i\sin\left(ka\right)}\right) & e^{ika}\frac{\tilde{V}}{2i\sin\left(ka\right)}\\
-e^{-ika}\frac{\tilde{V}}{2i\sin\left(ka\right)} & e^{-ika}\left(1-\frac{\tilde{V}}{2i\sin\left(ka\right)}\right)
\end{pmatrix}\begin{pmatrix}e^{ika\left(m-1\right)} & 0\\
0 & e^{-ika\left(m-1\right)}
\end{pmatrix}\begin{pmatrix}e^{ika}\left(1+\frac{\tilde{V}}{2i\sin\left(ka\right)}\right) & e^{ika}\frac{\tilde{V}}{2i\sin\left(ka\right)}\\
-e^{-ika}\frac{\tilde{V}}{2i\sin\left(ka\right)} & e^{-ika}\left(1-\frac{\tilde{V}}{2i\sin\left(ka\right)}\right)
\end{pmatrix}
\end{align}
with the elements of $\boldsymbol{M}$ simplifying to 
\begin{align}
M_{11} & =e^{ika\left(m+1\right)}\left(1+\frac{\tilde{V}}{2i\sin\left(ka\right)}\right)^{2}-e^{-ika\left(m-1\right)}\left(\frac{\tilde{V}}{2i\sin\left(ka\right)}\right)^{2}\nonumber \\
M_{22} & =e^{-ika\left(m+1\right)}\left(1-\frac{\tilde{V}}{2i\sin\left(ka\right)}\right)^{2}-e^{ika\left(m-1\right)}\left(\frac{\tilde{V}}{2i\sin\left(ka\right)}\right)^{2}=M_{\left(1,1\right)}^{*}\nonumber \\
M_{12} & =\frac{\tilde{V}}{2i\sin\left(ka\right)}\left[e^{ika\left(m+1\right)}\left(1+\frac{\tilde{V}}{2i\sin\left(ka\right)}\right)+e^{-ika\left(m-1\right)}\left(1-\frac{\tilde{V}}{2i\sin\left(ka\right)}\right)\right]\nonumber \\
M_{21} & =\frac{-\tilde{V}}{2i\sin\left(ka\right)}\left[e^{-ika\left(m+1\right)}\left(1-\frac{\tilde{V}}{2i\sin\left(ka\right)}\right)+e^{ika\left(m-1\right)}\left(1+\frac{\tilde{V}}{2i\sin\left(ka\right)}\right)\right]=M_{\left(1,2\right)}^{*}.\label{eq:dimer M elements}
\end{align}
From this transfer matrix we calculate 
\begin{align}
\left|M_{22}\right|^{2} & =M_{11}\times M_{22}\nonumber \\
 & =\left[e^{ika\left(m+1\right)}\left(1+\frac{\tilde{V}}{2i\sin\left(ka\right)}\right)^{2}-e^{-ika\left(m-1\right)}\left(\frac{\tilde{V}}{2i\sin\left(ka\right)}\right)^{2}\right]\nonumber \\
 & \,\,\,\times\left[e^{-ika\left(m+1\right)}\left(1-\frac{\tilde{V}}{2i\sin\left(ka\right)}\right)^{2}-e^{ika\left(m-1\right)}\left(\frac{\tilde{V}}{2i\sin\left(ka\right)}\right)^{2}\right]\nonumber \\
 & =\frac{1}{\sin^{2}\left(ka\right)}\left[\sin^{2}\left(ka\right)+\tilde{V}^{2}\left(\cos\left(kam\right)+\frac{\tilde{V}\sin\left(kam\right)}{2\sin\left(ka\right)}\right)^{2}\right],
\end{align}
\end{widetext}

which gives the transmission probability for a dimer of length $m$
as\citep{kim0673} 
\begin{equation}
T=\frac{\sin^{2}\left(ka\right)}{\sin^{2}\left(ka\right)+\tilde{V}^{2}\left[\cos\left(kam\right)+\frac{\tilde{V}\sin\left(kam\right)}{2\sin\left(ka\right)}\right]^{2}}.
\end{equation}
We now extend this calculation to a series of $N$ equal-strength
impurities, each separated by $m-1$ lattice sites.

For simplicity we define the parameters 
\begin{align}
g & =e^{ika}\\
v_{k} & =\frac{\tilde{V}}{2\sin\left(ka\right)}
\end{align}
so that the transfer matrix for a single impurity has the form 
\begin{equation}
\boldsymbol{M}_{1}=\begin{pmatrix}g\left(1-iv_{k}\right) & -igv_{k}\\
ig^{*}v_{k} & g^{*}\left(1+iv_{k}\right)
\end{pmatrix},
\end{equation}
and 
\begin{equation}
\boldsymbol{M}^{\left(m\right)}=\begin{pmatrix}g^{m} & 0\\
0 & g^{*m}
\end{pmatrix}.
\end{equation}
For $N$ impurities, the transfer matrix for the system is $\boldsymbol{M}=\boldsymbol{M}_{1}\boldsymbol{M}^{\left(m-1\right)}\boldsymbol{M}_{2}\cdots\boldsymbol{M}^{\left(m-1\right)}\boldsymbol{M}_{N}$,
where $\boldsymbol{M}_{1}=\boldsymbol{M}_{2}=\cdots=\boldsymbol{M}_{N}$
for equal-strength impurities. We therefore need to calculate the
matrix product $\boldsymbol{M}=\boldsymbol{M}_{1}\left(\text{\ensuremath{\boldsymbol{M}^{\left(m-1\right)}\boldsymbol{M}_{1}}}\right)^{\left(N-1\right)}$.
For a unimodular matrix $\boldsymbol{A}$, $\det\boldsymbol{A}=1,$
the Chebyshev identity\citep{mathworld} gives us 
\begin{align*}
\boldsymbol{A}^{N} & =\begin{pmatrix}A_{11} & A_{12}\\
A_{21} & A_{22}
\end{pmatrix}^{N}\\
 & =\begin{pmatrix}A_{11}U_{N-1}\left(h\right)-U_{N-2}\left(h\right) & A_{12}U_{N-1}\left(h\right)\\
A_{21}U_{N-1}\left(h\right) & A_{22}U_{N-1}\left(h\right)-U_{N-2}\left(h\right)
\end{pmatrix}
\end{align*}
where 
\begin{align}
h & =\frac{1}{2}\left(A_{11}+A_{22}\right)
\end{align}
and $U_{N}$ is the Chebyshev polynomial of the second kind. For $\text{\ensuremath{\boldsymbol{M}^{\left(m-1\right)}\boldsymbol{M}_{1}}}$,

\begin{align}
\text{\ensuremath{\boldsymbol{M}^{\left(m-1\right)}\boldsymbol{M}_{1}}} & =\begin{pmatrix}g^{\left(m-1\right)} & 0\\
0 & g^{*\left(m-1\right)}
\end{pmatrix}\begin{pmatrix}g\left(1-iv_{k}\right) & -igv_{k}\\
ig^{*}v_{k} & g^{*}\left(1+iv_{k}\right)
\end{pmatrix}\nonumber \\
 & =\begin{pmatrix}g^{m}\left(1-iv_{k}\right) & -ig^{m}v_{k}\\
ig^{*m}v_{k} & g^{*m}\left(1+iv_{k}\right)
\end{pmatrix}
\end{align}
and 
\begin{align}
h_{k} & =\frac{1}{2}\left(M_{11}+M_{22}\right)\nonumber \\
 & =\frac{1}{2}\left(g^{m}\left(1-iv_{k}\right)+g^{*m}\left(1+iv_{k}\right)\right)\nonumber \\
 & =\frac{1}{2}\left(g^{m}+g^{*m}-iv_{k}\left(g^{m}-g^{*m}\right)\right)\nonumber \\
 & =\frac{1}{2}\left(e^{ikam}+e^{-ikam}-iv_{k}\left(e^{ikam}-e^{-ikam}\right)\right)\nonumber \\
 & =\cos\left(kam\right)+v_{k}\sin\left(kam\right).
\end{align}

\begin{widetext}
Calculating the $(N-1)^{th}$ power gives us 
\begin{equation}
\left(\text{\ensuremath{\boldsymbol{M}^{\left(m-1\right)}\boldsymbol{M}_{1}}}\right)^{N-1}=\begin{pmatrix}g^{m}\left(1-iv_{k}\right)U_{N-2}\left(h_{k}\right)-U_{N-3}\left(h_{k}\right) & -ig^{m}v_{k}U_{N-2}\left(h_{k}\right)\\
ig^{*m}v_{k}U_{N-2}\left(h_{k}\right) & g^{*m}\left(1+iv_{k}\right)U_{N-2}\left(h_{k}\right)-U_{N-3}\left(h_{k}\right).
\end{pmatrix}
\end{equation}
The transfer matrix $\boldsymbol{M}$ for $N$ potentials, each separated
by $m$ lattice sites is 
\begin{align}
\boldsymbol{M} & =\boldsymbol{M}_{1}\left(\text{\ensuremath{\boldsymbol{M}^{\left(m-1\right)}\boldsymbol{M}_{1}}}\right)^{N-1}\nonumber \\
 & =\begin{pmatrix}g\left(1-iv_{k}\right) & -igv_{k}\\
ig^{*}v_{k} & g^{*}\left(1+iv_{k}\right)
\end{pmatrix}\begin{pmatrix}g^{m}\left(1-iv_{k}\right)U_{N-2}\left(h_{k}\right)-U_{N-3}\left(h_{k}\right) & -ig^{m}v_{k}U_{N-2}\left(h_{k}\right)\\
ig^{*m}v_{k}U_{N-2}\left(h_{k}\right) & g^{*m}\left(1+iv_{k}\right)U_{N-2}\left(h_{k}\right)-U_{N-3}\left(h_{k}\right)
\end{pmatrix}.\label{eq:M matrix general N}
\end{align}
For the transmission probability we only need to calculate $M_{11}$
from the above expression since $T=1/\left|M_{22}\right|^{2}=1/\left|M_{11}\right|^{2}$.
Following from Eq. (\ref{eq:M matrix general N}) 
\begin{align}
M_{11} & =g\left(1-iv_{k}\right)g^{m}\left(1-iv_{k}\right)U_{N-2}\left(h_{k}\right)-g\left(1-iv_{k}\right)U_{N-3}\left(h_{k}\right)+-igv_{k}ig^{*m}v_{k}U_{N-2}\left(h_{k}\right)\nonumber \\
 & =\left[g^{m+1}\left(1-iv_{k}\right)^{2}+g^{*\left(m-1\right)}v_{k}^{2}\right]U_{N-2}\left(h_{k}\right)-g\left(1-iv_{k}\right)U_{N-3}\left(h_{k}\right)
\end{align}
and 
\begin{align}
\left|M_{11}\right|^{2} & =\left(\left[g^{m+1}\left(1-iv_{k}\right)^{2}+g^{*\left(m-1\right)}v_{k}^{2}\right]U_{N-2}\left(h_{k}\right)-g\left(1-iv_{k}\right)U_{N-3}\left(h_{k}\right)\right)\nonumber \\
 & \times\left(\left[g^{*\left(m+1\right)}\left(1+iv_{k}\right)^{2}+g^{\left(m-1\right)}v_{k}^{2}\right]U_{N-2}\left(h_{k}\right)-g^{*}\left(1+iv_{k}\right)U_{N-3}\left(h_{k}\right)\right)\nonumber \\
 & =\left[g^{m+1}\left(1-iv_{k}\right)^{2}+g^{*\left(m-1\right)}v_{k}^{2}\right]\left[g^{*\left(m+1\right)}\left(1+iv_{k}\right)^{2}+g^{\left(m-1\right)}v_{k}^{2}\right]U_{N-2}^{2}\left(h_{k}\right)\nonumber \\
 & -g^{*}\left(1+iv_{k}\right)\left[g^{m+1}\left(1-iv_{k}\right)^{2}+g^{*\left(m-1\right)}v_{k}^{2}\right]U_{N-2}\left(h_{k}\right)U_{N-3}\left(h_{k}\right)\nonumber \\
 & -g\left(1-iv_{k}\right)\left[g^{*\left(m+1\right)}\left(1+iv_{k}\right)^{2}+g^{\left(m-1\right)}v_{k}^{2}\right]U_{N-2}\left(h_{k}\right)U_{N-3}\left(h_{k}\right)\nonumber \\
 & +g\left(1-iv_{k}\right)g^{*}\left(1+iv_{k}\right)U_{N-3}^{2}\left(h_{k}\right)
\end{align}
Simplifying all these terms yields 
\begin{align}
\left|M_{11}\right|^{2} & =U_{N-2}^{2}\left(h_{k}\right)\left(1+4h_{k}^{2}v_{k}^{2}\right)-U_{N-2}\left(h_{k}\right)U_{N-3}\left(h_{k}\right)2h_{k}\left(1+2v_{k}^{2}\right)+U_{N-3}^{2}\left(h_{k}\right)\left(1+v_{k}^{2}\right)
\end{align}
One can further simplify this equation above by using the recursion
relations for $U_{N}$, 
\begin{equation}
U_{N+1}\left(x\right)=2xU_{N}\left(x\right)-U_{N-1}\left(x\right),\label{recursion}
\end{equation}
and the expression that (proof in the following section) 
\begin{equation}
U_{N-2}^{2}\left(x\right)-2xU_{N-2}\left(x\right)U_{N-3}\left(x\right)+U_{N-3}^{2}\left(x\right)=1,
\end{equation}
one then obtains 
\begin{align}
\left|M_{11}\right|^{2} & =U_{N-2}^{2}\left(h_{k}\right)\left(1+4h_{k}^{2}v_{k}^{2}\right)-U_{N-2}\left(h_{k}\right)U_{N-3}\left(h_{k}\right)2h_{k}\left(1+2v_{k}^{2}\right)+U_{N-3}^{2}\left(h_{k}\right)\left(1+v_{k}^{2}\right)\nonumber \\
 & =1+v_{k}^{2}U_{N-1}^{2}\left(h_{k}\right).
\end{align}
\end{widetext}

The general expression for the transmission though $N$ potentials,
each separated by $m$ lattice sites is therefore\citep{griffiths01,markos08}
\begin{align}
T & =\frac{1}{1+v_{k}^{2}U_{N-1}^{2}\left(h\right)},
\end{align}
where $U_{N}\left(h\right)$ is the $N^{th}$ Chebyshev polynomial,
and 
\begin{align}
h & =\cos\left(kam\right)+v_{k}\sin\left(kam\right)\\
v_{k} & =\frac{V}{2t_{0}\sin\left(ka\right)}.
\end{align}

\section{Proof of $U_{N-2}^{2}-2hU_{N-2}U_{N-3}+U_{N-3}^{2}=1$ for all $N$\label{A:Proof}}

The recursion relation for Chebyshev polynomials of the second kind
is 
\begin{align}
U_{0}\left(h\right) & =1\nonumber \\
U_{1}\left(h\right) & =2h\nonumber \\
U_{2}\left(h\right) & =4h^{2}-1\nonumber \\
U_{3}\left(h\right) & =8h^{3}-4h\nonumber \\
U_{4}\left(h\right) & =16h^{4}-12h^{2}+1\nonumber \\
U_{n}\left(h\right) & =2hU_{n-1}\left(h\right)-U_{n-2}\left(h\right).\label{chebs}
\end{align}
Let $k=N-2$, we define 
\begin{equation}
d\left(k\right)\equiv U_{k}^{2}-2hU_{k}U_{k-1}+U_{k-1}^{2}.\label{dofk1}
\end{equation}
It is easy to show that $d(1)=1$ by direct substitution of Eq.~(\ref{chebs}).
Then judicious application of the recurrence relation for $U_{k+1}$
results in 
\begin{widetext}
\begin{eqnarray}
d\left(k+1\right) & = & U_{k+1}^{2}-2hU_{k+1}U_{k+1-1}+U_{k+1-1}^{2}\nonumber \\
 & = & \left(2hU_{k}-U_{k-1}\right)^{2}-2h\left(2hU_{k}-U_{k-1}\right)U_{k}+U_{k}^{2}\nonumber \\
 & = & \left(4h^{2}U_{k}^{2}-4hU_{k}U_{k-1}+U_{k-1}^{2}\right)-2h\left(2hU_{k}^{2}-U_{k}U_{k-1}\right)+U_{k}^{2}\nonumber \\
 & = & 4h^{2}U_{k}^{2}-4hU_{k}U_{k-1}+U_{k-1}^{2}-4h^{2}U_{k}^{2}+2hU_{k}U_{k-1}+U_{k}^{2}\nonumber \\
 & = & U_{k-1}^{2}-2hU_{k}U_{k-1}+U_{k}^{2}\nonumber \\
 & = & d\left(k\right)\label{dofk2}
\end{eqnarray}
\end{widetext}
for any $k$. If $d(1)=1,$ and $d(k)=d(k+1)$ for all $k$, then
$d(k)=1$ for all $k$, and the required identity is proven.

\renewcommand{\bibname}{References}


\begin{thebibliography}{10}

\bibitem{geiger1913} \bluee{H. Geiger and E. Marsden,  \textit{LXI, The laws of deflexion of $\alpha$-particles through large angles},
The London, Edinburgh, and Dublin Philosophical Magazine and Journal of Science, {\bf 25}, 604-623 (1913).}

\bibitem{kim0673} W. Kim, L. Covaci and F. Marsiglio, ``\textit{Hidden
symmetries of electronic transport in a disordered one-dimensional
lattice,}'' Phys. Rev. B \textbf{73}, 195109-1-5 (2006).

\bibitem{kim0674} Wonkee Kim, L. Covaci, and F. Marsiglio, ``{Impurity
scattering of wave packets on a lattice,}'' Phys. Rev. B \textbf{74},
205120-1-9 (2006).

\bibitem{norsen08} An excellent discussion of this very issue can
be found in T. Norsen, J. Lande and S.B. McKagan, ``\textit{How and
why to think about scattering in terms of wave packets instead of
plane waves,}'' arXiv:0808.3566 (2008). Parts of this manuscript
were published in T. Norsen, ``\textit{The pilot-wave perspective
on quantum scattering an tunnellng,}'' Am. J. Phys. \textbf{81},
258-266 (2013).

\bibitem{schonhammer19} K. Sch\"onhammer, ``\textit{Unusual broadening
of wave packets on lattices},'' Am. J. Phys. \textbf{87}, 186-193
(2019).

\bibitem{staelens21} M. Staelens and F. Marsiglio, ``\textit{Scattering
problems via real-time wave packet scattering},'' Am. J. Phys. \textbf{89},
693-701 (2021).

\bibitem{phet} See, for example,\\
 https://phet.colorado.edu/en/simulations/category/new.

%
%
%
%
%
%
%

\bibitem{phillips03} Philip Phillips, \textit{Advanced Solid State
Physics}, (Westview Press, Boulder, CO, 2003).

\bibitem{markos08} Peter Marko\v s and Costas M. Soukoulis, \textit{Wave
Propagation: From Electrons to Photonic Crystals and Left-Handed Materials},
(2nd ed. Princeton University Press, Princeton, 2008).

\bibitem{joannopoulos08} John D. Joannopoulos, Steven G. Johnson,
Joshua N. Winn and Robert D. Meade, \textit{Photonic Crystals: Molding
the Flow of Light}, (2nd ed. Princeton University Press, Princeton,
2008).

\bibitem{griffiths92} David J. Griffiths and Nicholas F. Taussig,
``\textit{Scattering from a locally periodic potential,}'' Am. J.
Phys. \textbf{60}, 883-888 (1992).

\bibitem{sprung93} D.W.L. Sprung, Hua Wu and J. Martorell, ``\textit{Scattering
by a finite periodic potential},'' Am. J. Phys. \textbf{61}, 1118-1124
(1993).

\bibitem{griffiths01} David J. Griffiths and Carl A. Steinke, ``\textit{Waves
in locally periodic media},'' Am. J. Phys. \textbf{69}, 137-154 (2001).
See Ref.~13 in this paper for earlier references, each of which carried
out some form of the same derivation in the continuum limit.

\bibitem{marsiglio09} \bluee{F.~Marsiglio, ``The harmonic oscillator in quantum mechanics: A third way,'' Am. J. Phys. \textbf{77}, 253--258 (2009).}

\bibitem{cmp_remark} \bluee{In fact there are many aspects of this problem that can be utilized in a course setting, particularly one that emphasizes topics
in condensed matter physics. For example, the eigenvalues for such a problem, given analytically in Appendix B, indicate $E_{k_0} = 0$ for $k_0 a = \pi/2$,
yet the wave packet moves, because it is the group velocity that plays the important role of governing the speed of the wave packet.}

\bibitem{dunlap90} D.H. Dunlap, H.-L. Wu and P.W. Phillips, ``\textit{Absence
of localization in a random-dimer model,}'' Phys. Rev. Lett. \textbf{65},
88-91 (1990).

\bibitem{wu91} H.-L. Wu and P.W. Phillips, ``\textit{Polyaniline
is a random-dimer model - a new transport mechanism for conducting
polymers,}'' Phys. Rev. Lett. \textbf{66}, 1366-1369 (1991).

\bibitem{wu92} H.-L. Wu, W. Goff and P.W. Phillips, ``\textit{Insulator-metal
transitions in random lattices containing symmetrical defects,}''
Phys. Rev. B \textbf{45}, 1623-1628 (1992).

\bibitem{datta93} P. K. Datta, D. Giri, and K. Kundu, ``\textit{Nonscattered
states in a random-dimer model,}'' Phys. Rev. B \textbf{47}, 10727-10737
(1993).

\bibitem{giri93} D. Giri, P. K. Datta, and K. Kundu, ``\textit{Tuning
of resonances in the generalized random trimer model,}'' Phys. Rev.
B \textbf{48}, 14113-14120 (1993).

\bibitem{mathworld} \href{https://mathworld.wolfram.com/UnimodularMatrix.html}{https://mathworld.wolfram.com/UnimodularMatrix.html}.

\end{thebibliography}
\end{document}